\documentclass[twocolumn,showpacs,amsmath,amssymb,prc,superscriptaddress,floatfix]{revtex4}
\usepackage{amsmath,amssymb}
\usepackage{graphicx}
\usepackage{graphics}
\usepackage{epsfig}

\newcommand{\dd}{\mbox{$\textrm{d}$}}
\newcommand{\half}{\mbox{${\textstyle \frac{1}{2}}$}}           

\begin{document}
\title{The extraction of parameters from final state interactions}

\author{G.~F\"{a}ldt}\email[E-mail: ]{goran.faldt@physics.uu.se}
\affiliation{Department of Physics and Astronomy, Uppsala University, Box 516, 751 20 Uppsala, Sweden}
\author{C.~Wilkin}\email[E-mail: ]{c.wilkin@ucl.ac.uk}
\affiliation{Physics and Astronomy Department, UCL, Gower Street, London WC1E
6BT, United Kingdom}

\date{\today}
\begin{abstract}
It is argued that final state enhancements in production reactions at large
momentum transfers, such as $pp\to K^+\Lambda p$, are primarily sensitive to
the position of a virtual bound state pole in the $\Lambda p$ system rather
than the $\Lambda p$ scattering length and effective range. These arguments
are supported by a study of the dispersion relation derived to describe such
processes as a function of the cut-off energy. This shows that the position
of the virtual bound state is independent of the cut-off energy.
\end{abstract}
\maketitle

\section{Introduction}
\label{Introduction}

It is difficult to study directly the scattering of some elementary particles
at low energies, especially if they are unstable and neutral, as is the case
for the $\Lambda$ hyperon. Nevertheless, a few low statistics measurements
have been made of the $\Lambda p$ cross sections in hydrogen bubble chambers
for center-of-mass (c.m.) momenta above about
60~MeV/$c$~\cite{ALE1968,SEC1968}.

An alternative approach is to look at the $\Lambda p$ system through a Final
State Interaction (FSI) in a production experiment, such as $pp\to K^+\Lambda
p$. By studying purely the kaon in an inclusive measurement, it is possible
to deduce directly the missing mass $m_X$ in the reaction and, for an
excitation energy in the $\Lambda p$ system $Q=(m_X-m_{\Lambda}-m_p)c^2 <
77$~MeV, conservation laws of strong interactions ensure that the unobserved
system $X=\Lambda + p$. The direct fitting of enhancements in such
data~\cite{SIE1994,BUD2010}, as described in Sec.~\ref{Direct}, shows that
there is a pole in the $\Lambda p$ scattering amplitude that corresponds to a
virtual (antibound) state in this system. However, the general pursuit in the
literature has not been into the extraction of the properties of this pole but
rather aimed at the determination of the $\Lambda p$ scattering length ($a$)
and the effective range ($r$), which depend in a model-dependent way also on
the behavior at much higher values of $Q$.

In an effort to avoid the high $Q$ problem, where the assumption that the
data are dominated by $S$ waves in the $\Lambda p$ system is highly
questionable, analyticity of the amplitudes has been exploited to give
estimates in terms of a dispersion integral over a finite range in
$Q$~\cite{GAS2004,GAS2005}. This method, which is described in Sec.~\ref{DR},
provides theoretical estimates of the uncertainties arising from the cut in
$Q$ and, to a lesser extent, those associated with the meson production
operator. It has been used extensively to extract information from exclusive
$pp\to K^+\Lambda p$ data taken by the COSY-TOF
collaboration~\cite{HAU2014,HAU2016}. However, once again the stress has been
placed on the determination of the scattering length~\cite{GAS2004,GAS2005}.

The numerical evaluations of the dispersion integral presented in
Sec.~\ref{Numerical} are largely consistent with the findings reported in the
theoretical papers~\cite{GAS2004,GAS2005}. With a cut-off at $Q_{\rm max} =
40$~MeV, the scattering length is reduced by about 0.3~fm but with important
changes also in the effective range. The main point of the current paper is
to show that, although $a$ and $r$ depend significantly on $Q_{\rm max}$, the
position of the nearby virtual state pole remains fixed when $Q_{\rm max}$ is
reduced from infinity to 40~MeV. Any uncertainty in determining the pole
position arising from the choice of $Q_{\rm max}$ is negligible compared to
the experimental errors. It is mainly this feature of the $\Lambda p$
interaction that is determined by the final state enhancement rather than the
scattering length and effective range separately. The dependence of the pole
position on the structure of the meson production operator also appears to be
much weaker.

Though the arguments given in this paper are couched mainly in terms of the
$pp\to K^+\Lambda p$ reaction, the principal results are valid much more
widely. Final state interactions in high momentum transfer reactions are more
sensitive to the positions of nearby bound or virtual state poles in the
scattering amplitudes rather than to asymptotic observables, such as the
scattering length.

As summarized in the conclusions of Sec.~\ref{Conclusions}, we have
investigated the effects of truncating the dispersion integral expression for
the $S$-wave phase shift. It was already clear from the work reported in
Refs.~\cite{GAS2004,GAS2005} that the resulting scattering length would be
modified and that this had to be taken into account. However, it was not
evident from these works that the position of the nearby pole in the
scattering amplitude was stable to changes in the cut-off parameter.

It is clear from its derivation that in the dispersion integral approach it
is assumed that the differential cross section for a high momentum transfer
production reaction, such as $pp\to K^+\Lambda p$, is determined by the
absolute square of the $\Lambda p$ Jost function, i.e., that the meson
production operator is of zero range. Though it is quite plausible that
short-distance effects dominate, changing the range of the production
operator would lead one to deduce a different value of the scattering length.
However, the position of the $\Lambda p$ virtual state pole is little
affected by the range in the production operator, and this is also discussed
in Sec.~\ref{Conclusions}.

\section{Direct fitting of FSI enhancements}
\label{Direct}

Experiments to measure inclusive kaon production in proton-proton collisions
were carried out at Saclay~\cite{SIE1994} and more recently by the HIRES
collaboration at COSY~\cite{BUD2010}. They both showed an enormous
enhancement just above the $\Lambda p$ threshold that is associated with the
$\Lambda p$ final state interaction (FSI).

\begin{figure}[hbt]
\begin{center}
\includegraphics[width=1.0\columnwidth]{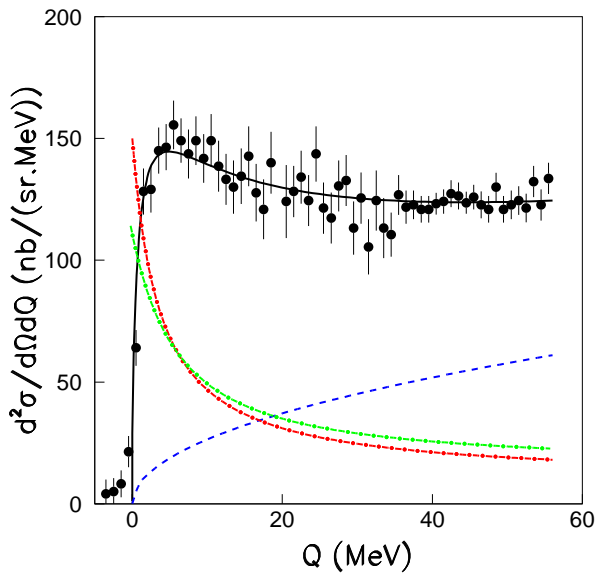}
\caption{\label{fig:HIRES4} (color online) Missing-mass spectrum of the
reaction $pp\to K^+\{X=\Lambda p\}$ measured at $\theta_K =0^{\circ}$ by the
COSY-HIRES collaboration at $T_p = 1.953$~GeV~\cite{BUD2010}. The (black)
solid line represents the description of Eq.~(\ref{directfit}), including the
$\Lambda p$ FSI generated by the Jost function of Eq.~(\ref{Jost}). The
experimental fit takes into account the missing-mass resolution with
$\sigma=0.84$~MeV, but this is not important for the present discussion and
is not shown. The (blue) dashed line represents the shape of the $pp\to
K^+\Lambda p$ phase space distribution, whereas the (red) chain curve is the
shape of the Jost representation of the $\Lambda p$ FSI. The arbitrarily
normalized FSI function of Eq.(\ref{analytic}) used in the analysis of the
COSY-TOF experimental data~\cite{HAU2016} is shown by the less steep (green)
chain curve.}
\end{center}
\end{figure}

The fit to the COSY-HIRES data in Fig.~\ref{fig:HIRES4} used the
factorization
\begin{equation}
\label{directfit}
\frac{\dd^2\sigma}{\dd\Omega_K\dd m_X} = A(m_X)\,\Phi_3 = |\mathcal{M}|^2\Phi_3|J(k)|^{-2},
\end{equation}
where $\Phi_3$ reflects the three-body phase space and $\mathcal{M}$ is a
production amplitude that is assumed to vary very slowly with the $\Lambda p$
relative momentum $k$. The crucial postulate is that the production operator
is of zero range such that the FSI can be expressed in terms of the Jost
function $J(k)$.

The COSY-HIRES data~\cite{BUD2010} are well described by a one-pole Jost
function $J(k)$, where
\begin{equation}
\label{Jost}
J(k) = \frac{k-i\alpha}{k+i\beta}\cdot
\end{equation}
The parameters deduced from the experimental data are $\alpha=(-0.31\pm
0.03)$~fm$^{-1}$ and $\beta=(1.21\pm 0.14)$~fm$^{-1}$, where the sign of
$\alpha$ is chosen such that there is no bound state in the $\Lambda p$
system and only statistical uncertainties are quoted. The Jost function of
Eq.~(\ref{Jost}) corresponds to the Bargmann potential~\cite{BAR1949}, for
which the effective range expansion
\begin{equation}
\label{ERE}
k\cot\delta(k) \approx -\frac{1}{a}+\half rk^2,
\end{equation}
is exact. Here $\delta(k)$ is the $S$-wave $\Lambda p$ phase shift. The
experimental results are~\cite{BUD2010}
\begin{equation}
a=\frac{\alpha+\beta}{\alpha\beta} = -2.43\pm 0.17~\textrm{fm},
 \ r=\frac{2}{\alpha+\beta} = 2.21\pm 0.16~\textrm{fm}.
\end{equation}
The immediate complication in the analysis is that there are two $\Lambda p$
$S$ waves, corresponding to the singlet and triplet combinations of the
hyperon and proton spins. The quoted parameters are therefore some average
over the two spin combinations, with unknown relative weights that depend on
the production mechanism and could vary with both the beam energy and the
kaon angle.

Although the data of Fig.~\ref{fig:HIRES4} are clearly sensitive to the
position of the nearby pole, i.e., the value of $\alpha$, the COSY-HIRES
authors stressed instead the values obtained for the scattering length and
effective range~\cite{BUD2010}, and this bias has continued in the more
sophisticated analysis discussed in the next section.

\section{Dispersion relation estimates}
\label{DR}

The direct fitting approach has been rightly criticized because it assumes
that, even at high excess energy $Q$, the $\Lambda p$ system remains in the
$S$ wave~\cite{GAS2004,GAS2005}. There are other complications at higher $Q$
because of the coupling to the $\Sigma N$ channel as well as distortions of
the matrix element $\mathcal{M}$ due to the excitation of $N^{\star}$
isobars. In an effort to avoid these problems, analyticity properties of the
production amplitude were exploited to yield a dispersion integral. Though
its full evaluation would require an integration over $Q$ to infinity, the
authors integrate only over a finite range, $0\leq Q \leq Q_{\rm max}$, and
then make theoretical estimates of the possible contributions from regions
beyond $Q_{\rm max}$~\cite{GAS2004,GAS2005}.

The starting point for the dispersion approach is similar to that of the
direct fitting procedure, which is that the $pp\to K^+\{\Lambda p\}$
double-differential cross section is proportional to the phase space
multiplied by the reciprocal of the absolute square of the $\Lambda p$ Jost
function. Having made this assumption, it is possible to generalize the
arguments given in the standard texts~\cite{GOL1964,NEW1982} to express the
$S$-wave phase shift in terms of two dispersion integrals. The integral up to
$Q_{\rm max}$ is over the logarithm of the Jost function and that beyond
$Q_{\rm max}$ depends upon the phase shift itself. In the approach adopted in
Refs.~\cite{GAS2004,GAS2005}, the phase shift is approximated by the first
finite integral.

In the cut-off approximation, the $S$-wave phase shift is given
by~\cite{GAS2004},
\begin{eqnarray}
\nonumber
\frac{\delta(k)}{k}&=&-\frac1{2\pi}\sqrt{\frac{m_{\rm min}}{m_{\rm red}}}\,
{\bf P} \int_{m_{\rm min}^2}^{m_{\rm max}^2}\dd\mu^2
\sqrt{\frac{m_{\rm max}^2-{m_X}^2}{m_{\rm max}^2-\mu^2}}\times\\
& & \hspace{-0.5cm}
\frac1{\sqrt{\mu^2-m_{\rm min}^2} \ (\mu^2-{m_X}^2)}
\log{\left\{A(\mu)\right\}}.
\label{final}
\end{eqnarray}
Here $m_{\rm red}$ is the reduced hyperon--proton mass and the minimum
missing mass is given by $m_{\rm min}=m_{p}+m_{\Lambda}$. In order to be
consistent with the potential description, the relation between $k$ and the
missing mass has been taken as the non-relativistic $m_X=m_{\rm min} +
k^2/2m_{\rm red}$. The choice of the cut-off parameter $m_{\rm max} = m_{\rm
min} +Q_{\rm max}/c^2$ is clearly subjective. Ideally it should be as large
as possible, subject to the $\Lambda p$ system still being in an $S$-wave. It
was argued that for $\Lambda$ production the best compromise would be
achieved by taking $Q_{\rm max}=40$~MeV~\cite{GAS2004}, which is well below
the $\Sigma N$ thresholds.

It should be noted that the integral vanishes if the argument of the
logarithm in Eq.~(\ref{final}) is constant so that, by subtracting the
integrand with $A(\mu)$ replaced by $A(m_X)$, it is possible to replace the
principal value (${\bf P}$) integral by a Riemann integral:
\begin{eqnarray}
\nonumber
\frac{\delta(k)}{k}&=&-\frac1{2\pi}\sqrt{\frac{m_{\rm min}}{m_{\rm red}}}\,
\int_{m_{\rm min}^2}^{m_{\rm max}^2}\dd\mu^2
\sqrt{\frac{m_{\rm max}^2-{m_X}^2}{m_{\rm max}^2-\mu^2}}\times\\
& & \hspace{-0.5cm}
\frac{1}{\sqrt{\mu^2-m_{\rm min}^2} \ (\mu^2-{m_X}^2)}
\log{\left\{\frac{A(\mu)}{A(m_X)}\right\}},
\label{final2}
\end{eqnarray}
where the singularity at $\mu = m_X$ in the denominator is canceled by the
zero from the logarithm.

The $k$ on the left hand side of Eq.~(\ref{final2}) is the $\Lambda p$ c.m.\
momentum corresponding to the missing mass $m_X$ so that the estimate of the
scattering length is obtained by taking the limit $m_X\to m_{\rm min}$, i.e.,
$a=-\lim_{k\to 0}\{\delta(k)/k\}$. However, by evaluating Eq.~(\ref{final2})
at a range of values of $k$ it is straightforward to extract approximations
for the effective range or higher terms in a $k^2$ expansion.

The numerical evaluation is facilitated if the integration variable is
changed to
\begin{equation}
\label{variable}
\mu^2 =m_{\rm min}^2\cos^2\theta +m_{\rm max}^2\sin^2\theta,
\end{equation}
which leads to
\begin{equation}
\frac{\delta(k)}{k}=-\frac{1}{\pi}\sqrt{\frac{m_{\rm min}(m_{\rm max}^2-{m_X}^2)}
{m_{\rm red}}}\!\!
\int_{0}^{\pi/2}\!\!\!\dd\theta\,
\frac{\log{\left\{A(\mu)/A(m_X)\right\}}}{(\mu^2-{m_X}^2)}\cdot
\label{final3}
\end{equation}
It should be noted that, if one takes the Jost function of Eq.~(\ref{Jost})
as input for $A$, then Eq.~(\ref{final3}) yields the correct Bargmann values
of the scattering length and effective range, but only in the limit where
$Q_{\rm max}\to \infty$. The pole is then at the ``correct'' place of
$k=i\alpha$.

The stability of the pole position is, however, a more general result. Below
threshold, $m_X < m_{\rm min}$, the dispersion denominator in
Eq.~(\ref{final3}) does not vanish and the integration over the
$\log\{A(\mu)\}$ term remains finite. There remains the second integral
proportional to $A(m_X)$ and, if this has a pole, as it does for example in
Eq.~(\ref{Jost}), then $\delta\to -i\infty$ at this point and the scattering
amplitude diverges. A virtual state pole would therefore remain in a fixed
position, independent of the cut-off.

\section{Numerical studies}
\label{Numerical}

\begin{figure}[hbt]
\begin{center}
\includegraphics[width=1.0\columnwidth]{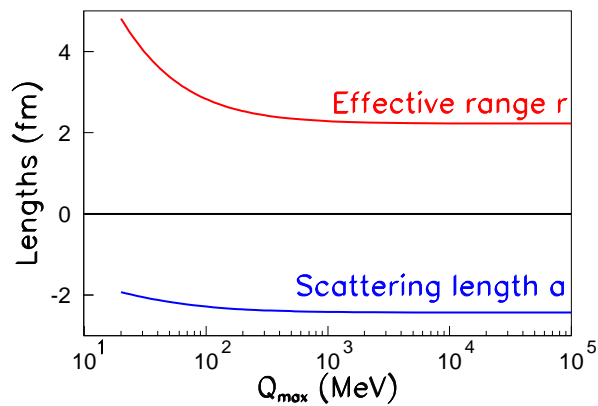}
\caption{\label{fig:gas7} (color online) Evaluations of the approximations to
the scattering length and effective range for the Jost function of
Eq.~(\ref{Jost}) as functions of $Q_{\rm max}$. The asymptotic parameters
used in this study, $\alpha=-0.307$~fm$^{-1}$ and $\beta=1.212$~fm$^{-1}$,
were those found by the COSY-HIRES collaboration~\cite{BUD2010}.}
\end{center}
\end{figure}

Values of the approximations to the scattering length and effective range for
the Jost function input of Eq.~(\ref{Jost}) were obtained through the
numerical evaluation of Eq.~(\ref{final2}) as functions of the cut-off energy
$Q_{\rm max}$. As shown in Fig.~\ref{fig:gas7}, for large $Q_{\rm max}$ these
approach the asymptotic values of $a=-2.43$ and $r=2.21$~\cite{BUD2010}. The
variation of the scattering length with $Q_{\rm max}$ is also consistent with
previous estimates~\cite{GAS2004}, changing by about 0.3~fm from its
asymptotic value by $Q_{\rm max} = 40$~MeV.

\begin{figure}[hbt]
\begin{center}
\includegraphics[width=1.0\columnwidth]{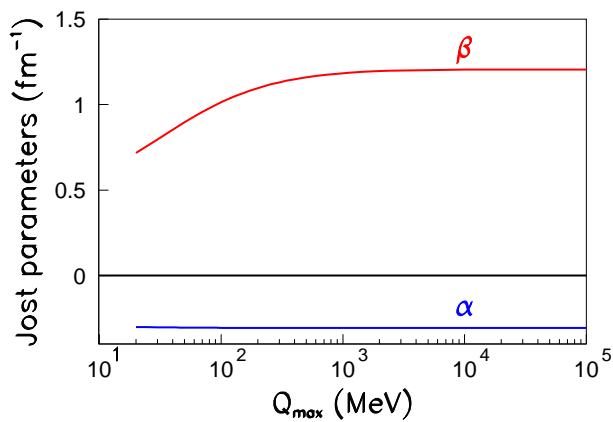}
\caption{\label{fig:gas8} (color online) Evaluations on the basis of the
truncated effective range expansion of Eq.~(\ref{ERE}) of the approximations
to the pole parameters for the Jost function of Eq.~(\ref{Jost}) as functions
of $Q_{\rm max}$. The asymptotic parameters used in this exercise,
$\alpha=-0.307$~fm$^{-1}$ and $\beta=1.212$~fm$^{-1}$, were those found by
the COSY-HIRES collaboration~\cite{BUD2010}. }
\end{center}
\end{figure}

Of much greater interest is the variation of the Jost parameters deduced by
fitting the effective range expansion of Eq.~(\ref{ERE}) to the values
extracted for the scattering length and effective range. Though, as shown  in
Fig.~\ref{fig:gas8}, at $Q_{\rm max}=40$~MeV $\beta$ deviates from its
asymptotic value by almost a third, the change in $\alpha$ is barely 1\%.
Variations of this size arise when finding the position of the nearby pole by
using the truncated effective range expansion of Eq.~(\ref{ERE}) and the
stability of $\alpha$ is confirmed by finding the pole position after keeping
the $k^4$ term in the effective range expansion. The result should also not
be understood as implying that the value of $\alpha$ is determined to 1\%
since the input function, which was not changed in the evaluation, has
significant experimental uncertainty. Rather the result signals that the
error in $\alpha$ induced by employing the truncated effective range
expansion with different choices of $Q_{\rm max}$ is negligible compared to
the other uncertainties.

If, instead of the Jost function of Eq.~(\ref{Jost}), one takes
\begin{equation}
\label{analytic}
A(m_X) \propto \exp[C_1^2/(m_X^2-C_2^2)],
\end{equation}
then the integral of Eq.~(\ref{final2}) can be evaluated to give an analytic
approximation for the scattering length~\cite{GAS2004} (and a more
complicated one for the effective range). This was the form assumed for the
extraction of the scattering length from the COSY-TOF data, with parameter
values of $C_1=348$~MeV/$c^2$ and $C_2=2039.2$~MeV/$c^2$~\cite{HAU2016}. It
must be noted that Eq.~(\ref{analytic}) should not be used below threshold
because it has the wrong analytic structure, with an essential singularity
rather than a simple pole. In such a case the stability of the virtual state
pole to changes in the cut-off energy cannot be proved a priori.

Nevertheless, as with the Jost input of Eq.~(\ref{Jost}), the estimates of
the scattering length and effective range change steadily with $Q_{\rm max}$
but the nearby pole, determined from the truncated effective range expansion,
remains far more stable, only changing from $\alpha=-0.421$~fm$^{-1}$ at
$Q_{\rm max}=100$~MeV to $\alpha=-0.417$~fm$^{-1}$ at $Q_{\rm max}=40$~MeV.
This value is somewhat different from the $\alpha=-0.31\pm0.04$~fm$^{-1}$
found by the COSY-HIRES collaboration~\cite{BUD2010} but systematic
uncertainties are hard to quantify. Any deviations between the two
experimental results, which seem to be already indicated by the FSI functions
shown in Fig.~\ref{fig:HIRES4}, could be due to disparities between the
spin-average COSY-TOF~\cite{HAU2014,HAU2016} and COSY-HIRES~\cite{BUD2010}
experimental data rather than the different methodologies in the fitting
processes. The larger value for $|\alpha|$ found at COSY-TOF is consistent
with the less steep enhancement factor shown in the curves of
Fig.~\ref{fig:HIRES4}.

\section{Conclusions}
\label{Conclusions}

The amplitude for a production reaction such as $pp\to K^+\Lambda p$ at large
momentum transfers is sensitive to the $\Lambda p$ wave function at short
distances. By taking the meson production operator to be of zero range, it is
assumed that the amplitude depends on the $\Lambda p$ Jost function. In the
standard approach, as adopted for example by the COSY-HIRES
collaboration~\cite{BUD2010}, the parameters of the Jost function are
obtained from a direct fit to the experimental data. Values of the scattering
length and effective range are then deduced from these parameters.

There are several problems in implementing such a direct fit approach. Since
the beam energy is finite, the data do not extend beyond some limited range
of excitation energy and other mechanisms are likely to become important at
large $Q$. Of immediate concern has to be the assumption that the $\Lambda p$
system remains in the $S$-wave for arbitrarily large values of $Q$. In an
attempt to avoid these problems, approximations for the $S$-wave phase shift
were obtained through a dispersion integral over a finite range of $0
\leqslant Q\leqslant Q_{\rm max}$ and estimates were made of the error in
truncating the integration at $Q_{\rm max}$~\cite{GAS2004,GAS2005}. It has
been argued that the best compromise for the $pp\to K^+\Lambda p$ reaction
would be to take $Q_{\rm max}=40$~MeV and this is the value used in the
analysis of the COSY-TOF data~\cite{HAU2014,HAU2016}.

The principal result of our analysis is that, even if one accepts all of the
assumptions made in Refs.~\cite{GAS2004,GAS2005}, it is clear that it is the
position of the nearby pole that is completely stable to changes in $Q_{\rm
max}$ rather than the value of the scattering length, which would therefore
require model-dependent corrections or error estimates. There are, of course,
other sources of error in the evaluation of the pole position but these are
largely independent of the choice of $Q_{\rm max}$.

The theoretical papers of Refs.~\cite{GAS2004,GAS2005} have pointed out that
it is possible to study the spin dependence of the $S$-wave $\Lambda p$
interaction through measurements of final state enhancements produced by
polarized proton beams. This important program has been pursued with the
COSY-TOF experimental data~\cite{HAU2014,HAU2016}. However, the COSY-HIRES
data were taken with an unpolarized beam in the forward direction where any
analyzing power would, in any case, vanish~\cite{BUD2010}. One can therefore
only compare the COSY-TOF and COSY-HIRES unpolarized data. Though these were
taken at similar beam energies, there could in principle still be differences
arising from the different angular coverage in the two experiments if the
$\Lambda p$ scattering parameters had a large spin dependence. Even if we
neglect such a possibility, the significance of the difference between the
values of $\alpha({\rm TOF})\approx -0.42$~fm$^{-1}$ and $\alpha({\rm
HIRES})= -0.31\pm0.04$~fm$^{-1}$ found in the two experiments must depend
upon a careful assessment of the systematic uncertainties.

The current understanding of the low energy $\Lambda$-nucleon interaction has
been nicely summarized in a recent review~\cite{GAL2016}. All the potentials
discussed there generate nearby $S$-wave virtual bound states in both the
spin-singlet and -triplet $\Lambda$-nucleon systems, with typically
$\alpha({\rm singlet}) \approx -0.28$~fm$^{-1}$ and $\alpha({\rm triplet})
\approx -0.38$~fm$^{-1}$. These are not very different to the values derived
from the COSY-HIRES and COSY-TOF data but these experimental results
correspond to spin averages with unknown weights.

Both the direct fitting and the dispersion relation approach assume that the
meson production operator in these high momentum transfer reactions is of
very short range. The validity of this assumption will, of course, depend on
both the form of the operator and the short-range $r$ behavior of the
two-body wave function. In the case of $pp\to \pi^+ pn$, the short-range
repulsion in the $pn$ system will diminish the importance of the behavior at
$r=0$~\cite{PAN1996}. To avoid this problem, the variation of the
Paris~\cite{LAC1980} $S$-wave spin-triplet $pn$ wave function with $k$ was
studied at $r=1.05$~fm~\cite{FAL1997}. Though the pole position, which is
fixed by the deuteron binding energy, is independent of the value chosen for
$r$, the momentum dependence of the wave function at $r=1.05$~fm is weaker
than that given by the Jost function. In the language of the Bargmann
potential, the value of $\beta$ would have been increased and so the
scattering length deduced from the Jost function would be in error, though
the value of $\alpha$ would be correct.

It was suggested several years ago that the positions of the nearby two-body
poles seem to govern the energy dependence of the total meson production
cross section near threshold~\cite{FAL1997a} and this approach has been used
to describe the $pp\to K^+\Lambda p$  total cross section at low
energies~\cite{SEW1999}. If the pole corresponds to a true bound rather than
a virtual one it is even possible to determine the normalization from the
residue at the pole. This allows one to estimate the total cross section for
$np$ spin-triplet production in $pp\to pn\pi^+$ in terms of that for $pp\to
d\pi^+$~\cite{FAL1996}.

We have largely ignored any questions of experimental errors and, with
perfect data, even a small range in $Q$ near threshold might be sufficient to
allow extrapolation to the pole. This is naturally a very naive limit but it
does offer a simple explanation as to why it is the value of $\alpha$
extracted from the dispersion integrals of Eq.~(\ref{final2}) that is almost
independent of $Q_{\rm max}$.

The inspiration of this work came from discussions with the authors of
Refs.~\cite{GAS2005}, \cite{HAU2016}, and \cite{BUD2010}. The constructive
suggestions made by A.~Gasparyan and J.~Haidenbauer were much appreciated.

%
%

\end{document}